# Reentrant ac magnetic susceptibility in Josephson-junction arrays: An alternative explanation for the paramagnetic Meissner effect

P. Barbara,* F. M. Araujo-Moreira,† A. B. Cawthorne, and C. J. Lobb
*Center for Superconductivity Research, Department of Physics, University of Maryland, College Park, Maryland 20742-4111*
(Received 16 January 1998; revised manuscript received 16 November 1998)

The paramagnetic Meissner effect (PME) measured in high-$T_C$ granular superconductors has been attributed to the presence of $\pi$ junctions between the grains. Here we present measurements of complex ac magnetic susceptibility from two-dimensional arrays of conventional (non-$\pi$) Nb/Al/AlOx/Nb Josephson junctions. We measured the susceptibility as a function of the temperature $T$, the ac amplitude of the excitation field $h_{ac}$ and the external magnetic field $H_{dc}$. The experiments show a strong paramagnetic contribution from the multi-junction loops, which manifests itself as a reentrant screening at low temperature, for values of $h_{ac}$ higher than 50 mOe. A highly simplified model, based on a single loop containing four junctions, accounts for this paramagnetic contribution and the range of parameters in which it appears. This model offers an alternative explanation of PME that does not involve $\pi$ junctions. [S0163-1829(99)02234-1]

## I. INTRODUCTION

An experimental study of the paramagnetic response in Bi-based polycrystalline superconductors was first reported by Braunisch et al.[1,2] They measured a paramagnetic dc susceptibility at values of temperature lower than the critical temperature $T_C$ of their superconducting samples. This paramagnetic response was in striking contrast to the usual diamagnetic Meissner effect, where magnetic field is excluded from superconductors. Although it is a misnomer, the anomalous response is widely referred to as paramagnetic Meissner effect (PME) in the literature; we will follow this custom in the rest of the paper.

The PME appeared systematically under specific experimental conditions and depended on sample preparation and morphology.

(a) The samples had to be cooled below $T_C$ in the presence of small magnetic field, $H < 1$ Oe; by increasing the value of $H$ in their field cooled (FC) experiments, they observed a crossover of the dc magnetic susceptibility to diamagnetic values.

(b) The PME was strongly dependent on the granular structure. Grinding the samples into small powder (a process that substantially weakens the contact between the grains) suppressed or even destroyed PME.

(c) Weak links with rather high critical currents were essential for the occurrence of PME: only melt-processed samples, more densely packed and with higher critical currents with respect to the sintered ones, showed PME.

Braunisch et al. also found that, after zero-field cooling (ZFC) the same samples, the measured susceptibility was diamagnetic. The authors attributed PME to the occurrence of spontaneous currents, flowing in direction opposite to ordinary Meissner screening currents. They proposed that anomalous Josephson junctions between the grains may be responsible for the existence of such currents. In these junctions ($\pi$ junctions) the Cooper pairs acquire a phase shift equal to $\pi$ in the tunneling process and the Josephson current has direction opposite to conventional junctions. $\pi$ junctions may be the consequence of magnetic impurities in the junction,[3,4] or non-$s$-wave pairing symmetry.[5]

Regardless of the origin of $\pi$ junctions, Dominguez et al.[6] have modeled a granular superconductor by considering a network where the nodes represent the grains and the links represent the coupling between the grains. This network was a mixture of normal junctions and $\pi$ junctions. In this case, the low-temperature and low-field ZFC susceptibility was of the order of $-1$ (in SI units), while the FC susceptibility was paramagnetic for some values of magnetic field, reproducing qualitatively the experimental data obtained from Bi-based superconductors.

Other models based on networks of conventional junctions could explain the PME experimental results. Auletta et al.[7] found that numerical simulations of a two-dimensional array of conventional Josephson junctions, made of concentric multijunction loops, lead to positive FC magnetic susceptibility, qualitatively similar to the experimental PME.

Our experiments on the ac magnetic susceptibility of two-dimensional Josephson junction arrays, in ZFC experiments, are in agreement with this last picture, i.e., they show that networks of conventional Josephson junctions can give a paramagnetic contribution to the measured susceptibility. We use a simple multijunction loop model to explain how, in spite of the very different experimental conditions, our experiment can provide an alternative explanation for PME.

We recently published part of our results in Ref. 8. Here, we give a more detailed description of this work, and we include more recent experiments and numerical simulations.

## II. EXPERIMENTAL RESULTS

Complex ac magnetic susceptibility is a powerful low-field technique that has been successfully used to measure properties such as critical temperature, critical current density, and penetration depth in superconductors. To measure samples in the shape of thin films, the so-called screening method has been developed. It involves the use of primary and secondary coils, with diameters smaller than the dimension of the sample. When these coils are located near the surface of the film, the response, i.e., the complex output voltage $V$, does not depend on the radius of the film or its properties near the edges. In the reflection technique,[9] an excitation coil (primary) coaxially surrounds a pair of counterwound pickup coils (secondaries). When there is no sample in the system, the net output from these secondary coils is close to zero, since the pickup coils are close to identical in shape, but are wound in opposite directions. The sample is positioned as close as possible to the set of coils, to maximize the induced signal on the pick up coils (see Fig. 1).

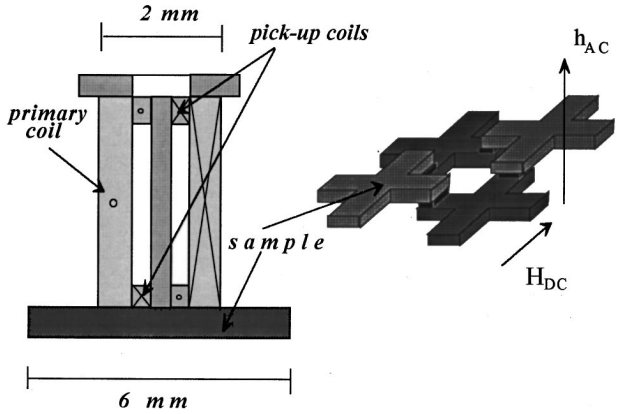

FIG. 1. Schematic of the experimental setup. The two pickup coils are counterwound. A unit cell of the array is shown in the inset: the crosses are niobium islands and the junctions are in the overlap region between them. $h_{ac}$ is the oscillating excitation field from the primary coil and $H_{dc}$ is a dc magnetic field, applied in the direction parallel to the sample.

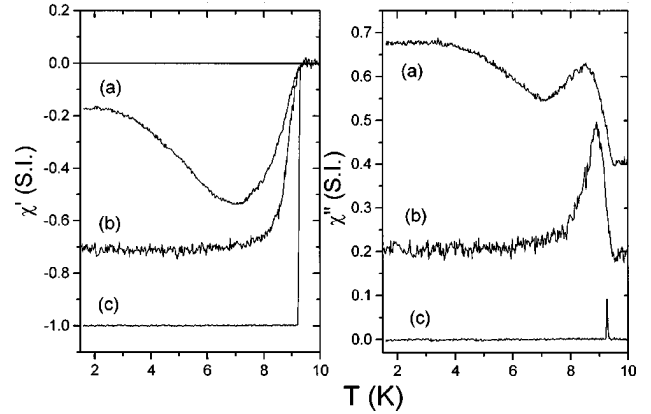

FIG. 2. $\chi'$ and $\chi''$ from the array as a function of $T$, for (a) $h_{ac}=96$ mOe, (b) 7 mOe, and (c) from a 500-nm thick niobium film for $h_{ac}=10$ mOe. The curves (a) and (b) for $\chi''$ have been vertically shifted by 0.2 (SI) for clarity.

An ac current is applied to the primary coil to create a magnetic field of amplitude $h_{ac}$ and frequency $f$. The output voltage of the secondary coils $V$ is a function of the complex susceptibility, $\chi_{ac}=\chi'+i\chi''$, and is measured through the usual lock-in technique. If we take the current on the primary as a reference, $V$ can be expressed by two orthogonal components, the inductive component, $V_L$ (in phase with the time derivative of the reference current), and the quadrature resistive component, $V_R$ (in phase with the reference current). This means that $V_L$ and $V_R$ are correlated with the average magnetic moment and the energy losses of the sample, respectively.

We used the screening method in the reflection configuration to measure $\chi_{ac}(T)$ of Josephson junction arrays. Measurements were performed as a function of the temperature $T$ (1.5 K $<T<$ 15 K), the amplitude of the excitation field $h_{ac}$ (1 mOe $<h_{ac}<$ 10 Oe), and the external magnetic field $H_{dc}$ (0 $<H_{dc}<$ 100 Oe) parallel with the plane of the sample. The frequency in the experiments reported here was fixed at $f=1.0$ kHz. The susceptometer was positioned inside a double-wall $\mu$-metal shield, screening the sample region from Earth's magnetic field. Our samples consisted of 100 $\times$150 unshunted tunnel junctions. The unit cell had square geometry with lattice spacing $a=46$ $\mu$m and the junction area of $5\times 5$ $\mu$m$^2$. From these dimensions, we estimated that the inductance of each loop was about 64 pH (see Fig. 1). The critical current density for the junctions forming the arrays was about 600 A/cm$^2$ at 4.2 K, giving $I_C\simeq 150$ $\mu$A for each junction.

We have performed four different types of experiments: (1) $\chi_{ac}(T)$, for different fixed values of $h_{ac}$, and no external parallel magnetic field, $H_{dc}$; (2) $\chi_{ac}(T)$, for a fixed value of $h_{ac}$, and different values of $H_{dc}$; (3) $\chi_{ac}(h_{ac})$, for fixed values of the temperature, and no external parallel magnetic field, $H_{dc}$; (4) $I_C(H_{dc})$ for a fixed temperature, from transport current-voltage characteristics.

In Fig. 2 we show results for $\chi_{ac}(T)$, obtained from ZFC experiments, for $h_{ac}=7$ and 96 mOe, and with $H_{dc}=0$. For $h_{ac}$ smaller than about 50 mOe, the behavior of both components of $\chi'(T)$ is quite similar to typical superconducting samples.[10] $\chi'(T)$, which is a measure of the screening current, becomes more negative at lower temperatures, indicating stronger superconductivity through the Meissner effect. $\chi''(T)$ peaks, indicating a maximum in the losses, around the critical temperature $T_C$. Notice that $\chi'\simeq -0.7(SI)$ for $h_{ac}=10$ mOe, at low temperature. The sample can only partially screen the external magnetic field.[11] As we can see in Fig. 2, the screening of the array is weaker compared to screening of a thick (500-nm) niobium film. We therefore define the array to be in a Meissner-like state for $h_{ac}<50$ mOe. This partial screening will be qualitatively explained in the following section, through the single-loop picture.

Outside the Meissner-like regime, for values of $h_{ac}>50$ mOe, $\chi'(T)$ is reentrant. It first increases in modulus as the temperature is lowered from the critical temperature $T_C$, then decreases at a lower temperature. The minimum in $\chi'(T)$ appears at $T\approx 7.0$ K. In all the temperature range, at a fixed $T$ the modulus of $\chi'$ decreases by increasing $h_{ac}$. The out-of-phase component, $\chi''(T)$, is correlated with the reentrance observed in $\chi'(T)$, showing increasing losses as the screening decreases, indicating an apparent weakening of the order parameter at low temperatures.

To experimentally investigate the origin of the reentrance, we have measured $\chi_{ac}(T)$ at a fixed value of the amplitude of the excitation field, $h_{ac}=96$ mOe, for different values of $H_{dc}$ (see Fig. 3). The external magnetic field $H_{dc}$ is parallel to the plane of the array (see Fig. 1). For our sample geometry, this parallel field suppresses the critical current $I_C$ of each junction, while inducing a negligible flux into the ''holes'' of the array (we estimate an alignment to the parallel direction within 0.1°). The measurements show that the position of the reentrance is being tuned by $H_{dc}$. We also observe that the value of temperature $T^*$ at which of $\chi'(T)$ has a minimum, shifts towards lower temperatures as we raise $H_{dc}$, down to $T^* \simeq 6$ K at $H_{dc}=33$ Oe. Further increase of $H_{dc}$ shifts $T^*$ back to higher temperature.

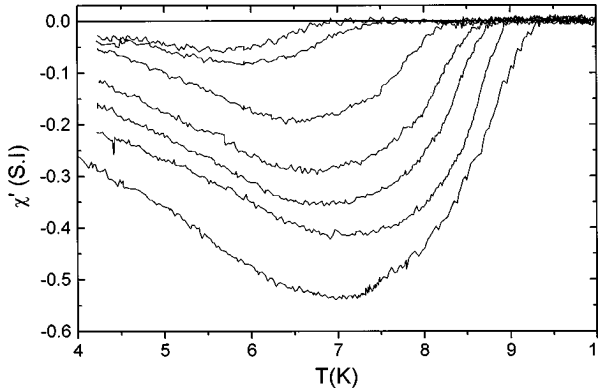

FIG. 3. $\chi'$ as a function of $T$ for different values of $H_{dc}$. From the bottom to the top $H_{dc}=0.0$, 6.5, 13.0, 19.5, 26.0, 30.5, and 32.5 Oe.

This nonmonotonic behavior is similar to the dependence of the Josephson junction critical current on a magnetic field applied in the plane of the junction[13] (Fraunhofer pattern). We measured $I_C(H_{dc})$ from transport current-voltage characteristics, at different values of $H_{dc}$ and at $T=4.2$ K (see Fig. 4). We find that $\chi'(T=4.2$ K), obtained from the isotherm $T=4.2$ K (see Fig. 3), shows the same Fraunhofer-like dependence on $H_{dc}$ as the critical current $I_C$ of the junctions forming the array. This gives further proof that only the junction critical current is varied in this experiment (Fig. 4). This also indicates that the screening currents at low temperature (i.e., the reentrant region) are proportional to the critical currents of the junctions. Furthermore, this shows an alternative way to obtain $I_C(H_{dc})$ in big arrays.

We have also determined the dependence of $\chi'$ and $\chi''$ on $h_{ac}$. This is an important technique for studying the critical state of superconducting materials.[14] Figure 5 shows the measured $\chi$ vs $h_{ac}$ in our arrays, at $T=4.2$ K. We observe that there is a sharp increase in both $\chi'$ and $\chi''$, around $h_{ac}$ $=50$ mOe. The amplitude of this jump decreases as we increase the temperature, until $T\approx 5.0$ K. For $T>5$ K there is no longer a discontinuity. This discontinuity in the curve is a signature of the transition from the Meissner-like regime to the reentrant regime, as we will show in the next section.

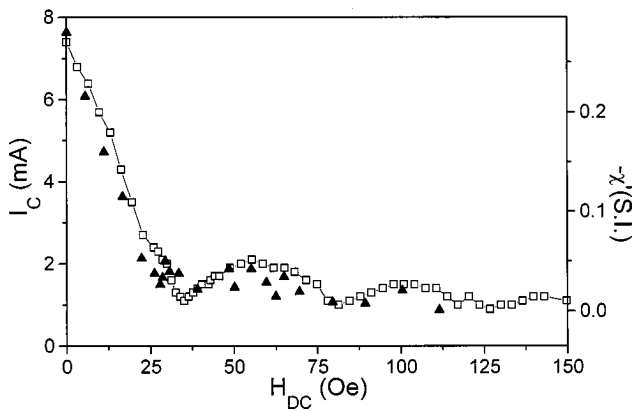

FIG. 4. $I_C$ (open squares) and $\chi'$ (solid triangles) as a function of $H_{dc}$ at $T=4.2$ K.

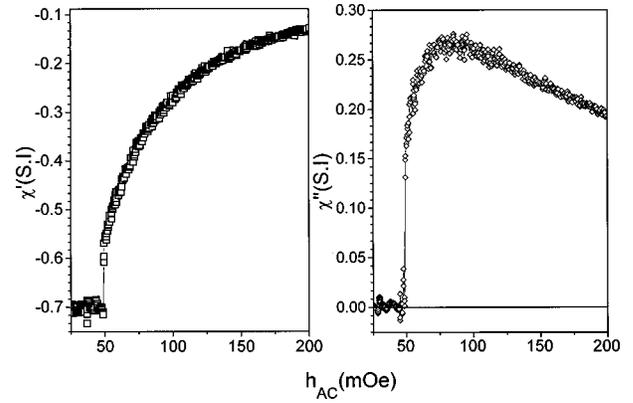

FIG. 5. $\chi'$ and $\chi''$ as a function of $h_{ac}$ at $T=4.2$ K.

## III. NUMERICAL RESULTS

We have found that all the experimental results can be qualitatively explained by analyzing the dynamics of a single unit cell in the array. The idea to use a single unit cell to qualitatively understand PME was first suggested by Auletta et al.[15] They simulated the field-cooled dc magnetic susceptibility (the same experimental conditions used by Ref. 1 to measure PME) of a single-junction loop and found a paramagnetic signal at low values of external magnetic field (a later extension of the model to a two-dimensional array[7] gave qualitatively the same results).

In our experiment, the unit cell is a loop containing four junctions and the measurements correspond to ZFC ac magnetic susceptibility. We will show that, notwithstanding our experimental conditions being different, the reentrant behavior in $\chi_{ac}$ is a signature of paramagnetic contributions from the multijunction loops, in agreement with the conclusions of Refs. 7 and 15.

We briefly outlined this model in Ref. 8. Here we will recall a few basic equations and we will show more numerical results.

We model a single unit cell as having four identical junctions, each with capacitance $C_J$, quasiparticle resistance $R_J$, and critical current $I_C$. We apply an external field of the form

$$H_{EXT}=h_{ac}\cos(\omega t). \quad (1)$$

The total magnetic flux threading the four-junction superconducting loop is given by

$$\Phi_{TOT}=\Phi_{EXT}+LI, \quad (2)$$

where $\Phi_{EXT}=\mu_0 a^2 H_{EXT}$, with $\mu_0$ being the vacuum permeability, $I$ is the circulating current in the loop, and $L$ is the inductance of the loop. The current $I$ is given by

$$I=I_c\sin\gamma_i+\frac{\Phi_0}{2\pi R_J}\frac{d\gamma_i}{dt}+\frac{C_J\Phi_0}{2\pi}\frac{d^2\gamma_i}{dt^2}, \quad (3)$$

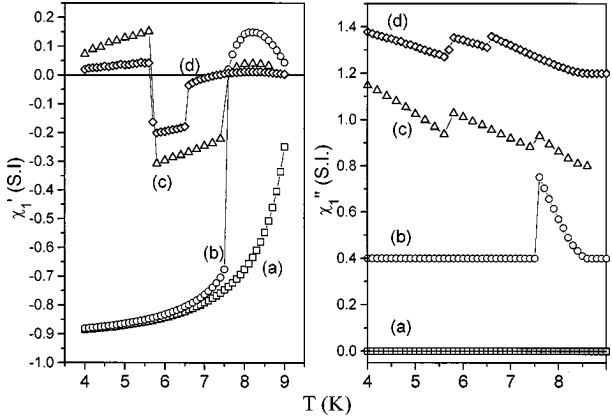

FIG. 6. Simulations of $\chi_1$ as a function of $T$ for (a) $h_{ac}$ = 5 mOe, (b) 29 mOe, (c) 69 mOe, and (d) 118 mOe. $\beta_L(4.2\,\mathrm{K})$ = 30 and $\beta_C(4.2\,\mathrm{K}=60)$. The curves for $\chi_1''$ have been vertically shifted by 0.4 (SI) for clarity.

where $\gamma_i$ is the the gauge-invariant superconducting phase difference across the $i$th junction and $\Phi_0$ is the magnetic flux quantum.

The magnetization

$$M = \frac{LI}{\mu_0 a^2} \quad (4)$$

may be expanded as a Fourier series

$$M(t) = h_{ac} \sum_{m=0}^{\infty} [\chi_m' \cos(m\omega t) + \chi_m'' \sin(m\omega t)]. \quad (5)$$

We calculated $\chi_1'$ and $\chi_1''$ through Eq. (5). Both Euler and fourth-order Runge-Kutta integration methods provided the same numerical results. In the model we do not include other effects (such as thermal activation) beyond the above equations.

In the model, the temperature-dependent parameter is the critical current of the junctions, given to good approximation by[16]

$$I_C(T) = I_C(0) \sqrt{1 - \frac{T}{T_C}} \tanh\left[1.54 \frac{T_c}{T} \sqrt{1 - \frac{T}{T_c}}\right]. \quad (6)$$

We calculated $\chi_1$ as a function of $T$. $\chi_1$ depends on the parameter $\beta_L$, which is proportional to the number of flux quanta that can be screened by the maximum critical current in the junctions, and the parameter $\beta_C$, which is proportional to the capacitance of the junction:

$$\beta_L(T) = \frac{2\pi L I_C(T)}{\Phi_0}, \quad (7)$$

$$\beta_C(T) = \frac{2\pi I_C(T) C_J R_J^2}{\Phi_0}. \quad (8)$$

We calculated $\beta_L(T=4.2\,\mathrm{K}) \simeq 30$ and $\beta_C(T=4.2\,\mathrm{K}) \simeq 60$.

The simulated $\chi_1'(T)$ and $\chi_1''(T)$, for different values of $h_{ac}$, are shown in Fig. 6. For values of $h_{ac}$ smaller than 47 mOe (corresponding to $\mu_0 a^2 h_{ac} \simeq 5\Phi_0$), $\chi_1'$ decreases with decreasing temperature, and $\chi_1''$ is close to zero.

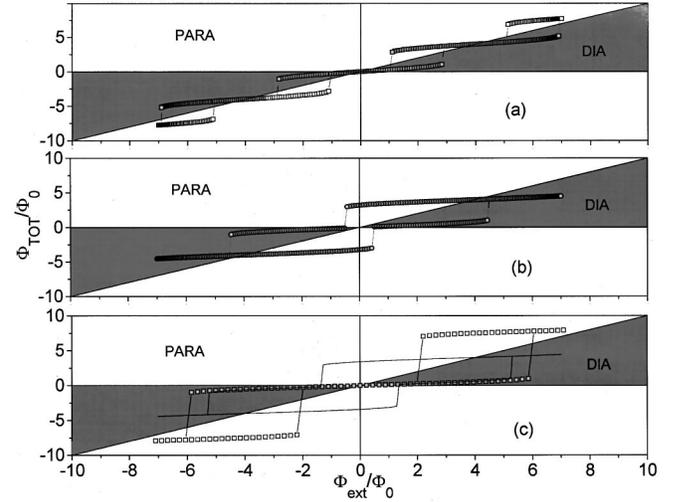

FIG. 7. Simulations of $\Phi_{TOT}$ as a function of $\Phi_{EXT}$ at (a) $T$ = 7.6 K, (b) 6.0 K, and (c) 4.0 K. $\beta_L(4.2\,\mathrm{K})=30$. $\beta_C(4.2\,\mathrm{K})=60$ (open squares) and $\beta_C(4.2\,\mathrm{K})=2$ (continuous solid line).

By increasing $h_{ac}$ above 47 mOe, reentrance at low temperature clearly appears and the screening becomes weaker in all the temperature range. This is consistent with the experiment, where the magnitude of $\chi'$ decreases with increasing $h_{ac}$. Note that at these high values of $h_{ac}$ the simulated $\chi_1''$ increases significantly, i.e., the simulation reproduces the dramatic increase of losses at the low temperature that is found in the experiment (see Fig. 2).

In the simulated $\chi_1'$ the reentrance appears as a paramagnetic region below 5 K. Another paramagnetic region is present above 7.5 K. These regions correspond to the measured decrease of screening below and above the minimum value of $\chi_1'$ in Fig. 2. The simulated $\chi_1'$ is either paramagnetic or diamagnetic, depending on the temperature range.

This surprising result can be understood by calculating the curves $\Phi_{TOT}$ vs $\Phi_{EXT}$, at different temperatures. As an example, in Fig. 7 we plot these curves for the same parameters used for curve (c) in Fig. 6, so $\Phi_{EXT} = \mu_0 h_{ac} a^2 \cos \omega t$ $= 7\Phi_0 \cos \omega t$. At low values of temperature, these curves are very hysteretic, showing multiple branches [see Fig. 7(c)]. The hysteresis decreases with increasing the temperature (and eventually disappears at $T \simeq 8.5\,\mathrm{K}$).

The important aspect of these curves is that they contain both paramagnetic and diamagnetic states. In Fig. 7, the line $\Phi_{TOT} = \Phi_{EXT}$ marks the boundary between diamagnetic states and paramagnetic states. For clarity, we shaded the diamagnetic areas in the graph ($\Phi_{TOT} < \Phi_{EXT}$), while clear areas correspond to paramagnetic states ($\Phi_{TOT} > \Phi_{EXT}$).

At a fixed value of temperature and $h_{ac}$, the value of $\chi_1$ is a time average of all the magnetic states that the system transverses during one cycle of $H_{EXT}$. In other words, $\chi_1$ is either diamagnetic or paramagnetic, depending on the shape of the part of the hysteresis curve that is spanned during one cycle of $H_{EXT}$.

The shape of the curve $\Phi_{TOT}$ vs $\Phi_{EXT}$ changes with temperature. For example, Fig. 7(a) at $T=7.6\,\mathrm{K}$ has three stable branches at positive values of $\Phi_{EXT}$. Note that one branch is completely paramagnetic. This causes the average response to be paramagnetic, making $\chi_1'$ positive. This scenario corresponds to the positive values of $\chi_1'$ at $T>7.5$ K, shown in Fig. 6(c).

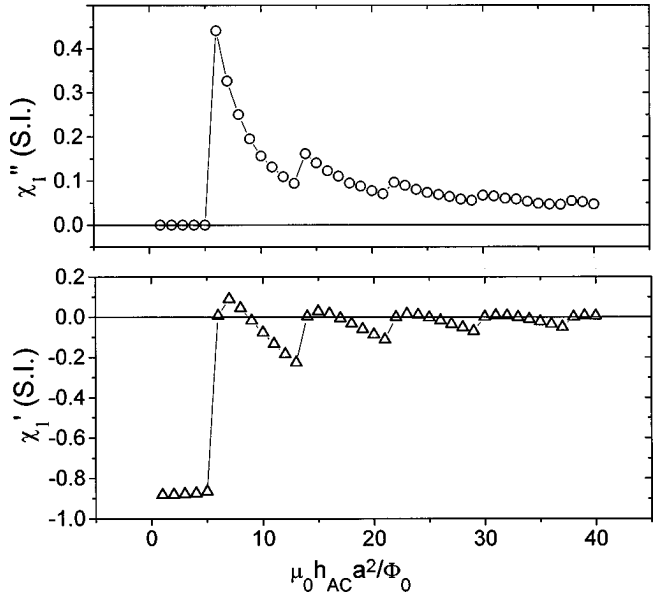

FIG. 8. Simulations of $\chi_1'$ and $\chi_1''$ as a function of $h_{ac}$ at $T=4.2$ K. $\beta_C=60$ and $\beta_L=30$.

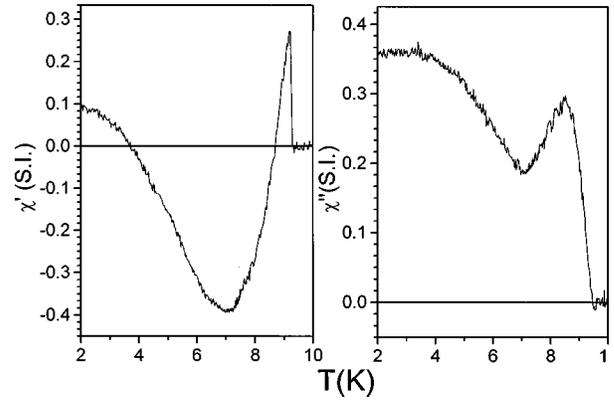

FIG. 9. Contribution to $\chi'$ and $\chi''$ vs $T$ from the superconducting loops at $h_{ac}=96$ mOe. These curves are obtained by subtracting the contribution of the niobium islands from the data in Fig. 2.

By lowering the temperature, the length of all the branches increases until, at about $T=7.5$ K, the second branch extends in the diamagnetic region up to the highest values of $\Phi_{EXT}$ ($7\Phi_0$) and the third branch becomes inaccessible. In fact, for $T<7.5$ K, the curve $\Phi_{TOT}$ vs $\Phi_{EXT}$ consists of two branches, as shown in Fig. 7(b). In this region, the average is diamagnetic, i.e., $\chi_1'$ is negative. (Note that the branch crossing $\Phi_{TOT}=0$ is all diamagnetic, while the second branch is diamagnetic at high values of $\Phi_{EXT}$ and paramagnetic at low values of $\Phi_{EXT}$: the average turns out to be diamagnetic.) This scenario corresponds to the negative values of $\chi_1'$ at $5<T<7.5$ K, shown in Fig. 6(c).

By further lowering the temperature, below $T\simeq 5$ K, the first branch extends to higher values of $\Phi_{EXT}$, where the third branch becomes stable and the second unstable (at $\Phi_{TOT}/\Phi_0=6$ there is a crossover of stability between the second and the third branch). Therefore, the system switches directly from the first branch to the third, which is fully paramagnetic [see Fig. 7(c)]. In this case, the average response is paramagnetic and $\chi_1'$ is positive. This scenario corresponds to the positive values of $\chi_1'$ at $T<5$ K, shown in Fig. 6(c).

Our simulations show that different values of $\beta_C$ affect the switching point (i.e., the length) of the branches in the low-temperature range. For example, we show that the curve at $T=4.2$ K for $\beta_C=2$ [solid line in Fig. 7(c)] is qualitatively very similar to the curve at $T=6.0$ K, Fig. 7(b). This happens because of the early switch from the first to higher branches occurring at lower values of $\Phi_{EXT}$, i.e., at values of $\Phi_{EXT}$ where the second branch is still stable. For small values of $\beta_C$, our simulations show no reentrance at low temperature.

Figure 8 shows the calculated $\chi_1$ as a function of $h_{ac}$ at $T=4.2$ K. The large jump in $\chi_1'$ corresponds to magnetic flux entering the loop at the switch from the first branch to higher branches. The calculated value of external field (47 mOe) corresponding to this jump is very close to the measured value (50 mOe), see Fig. 5.

In the simulated $\chi_{ac}$ the minimum value of $\chi'$ is about $-0.9$ (see Fig. 8), meaning that the sample can only provide partial screening. This is due to the fact that the diamagnetic branch crossing the value $\Phi_{EXT}=0$ (see Fig. 7) has a non-zero slope, i.e., some flux penetrates the sample even in the Meissner-like regime. What distinguishes the Meissner-like regime from the reentrant regime is the fact that the Meissner-like regime is reversible, while the reentrant regime is not. In fact, the reentrant regime involves switching to higher branches, which introduces pinning and hysteresis, in other words, irreversibility.

An analysis of $\chi_1''$ confirms this picture: the losses are negligible in the Meissner-like regime and increase significantly in the reentrant regime.

## IV. DISCUSSION

Surprisingly, numerical simulations of a very simple model, a four-junctions loop, account very satisfactorily for our experimental results, suggesting that this reentrance is dynamic in origin. However, we cannot make a completely quantitative comparison between our model and the measured array. The response of the array results from an average of the response from many loops. The flux distribution in the array is, in general, nonuniform, giving rise to different values of $h_{ac}$ in different loops. As a consequence, the measured response of the arrays presents no sharp transitions. The profile of the field penetration in the whole array has been analyzed by other authors,[17–24] but is not included in our model.

We can only make a quantitative comparison with our model in the Meissner-like state. In this case, the screening current in each loop is equivalent to a screening current flowing through a very large loop of junctions (about the dimension of the diameter of the coil).

In the reentrant state, it is instructive to compare the simulation with the measured response after subtracting the contribution of the superconducting islands. From our data, we can subtract the contribution of the niobium from the measured $\chi_{ac}$ of the array. In Fig. 9 we subtract the measured niobium response (multiplied by the factor 11/46, which accounts for the fraction of volume the niobium occupies in the sample) from the data corresponding to the curve of $\chi_{ac}$ vs $T$ at $h_{ac}=96$ mOe, which we showed in Fig. 2. The resulting curve shows paramagnetic response both at low and high temperature, analogous to the simulations in Fig. 6. Both the nonuniform flux distribution in the array and the contribution of the niobium islands are responsible for the fact that our total measured response is always diamagnetic. The signature of the paramagnetic contribution from some multijunction loops is the reentrance, but the paramagnetic contribution is never sufficiently strong to change the sign of the (total) measured $\chi_1'$.

## V. SUMMARY AND CONCLUSIONS

We measured PME from a network of conventional (non-$\pi$) Josephson junctions. In our experiments, PME occurs in the form of a low-temperature reentrance in the ac magnetic susceptibility. This reentrance appears for values of $h_{ac}$ higher than about 50 mOe, in excellent agreement with our estimated value of $LI_C/\mu_0 a^2 = 3.7$ A/m$=47$ mOe. This value of $h_{ac}$ is a threshold for the screening of the sample, i.e., for the Meissner-like state. Above this value, magnetic flux penetrates the sample and is pinned, because of the high value of $I_C$ and $\beta_L$. (High values of $\beta_L$ correspond to strong pinning, i.e., strong hysteresis in the $\Phi_{TOT}$ vs $\Phi_{EXT}$ curve.) Through numerical simulations of a simple model, we showed that the multijunction loops are paramagnetic in the reentrant region. Moreover, by subtracting the contribution of the niobium islands from the measured reentrant susceptibility, we find a similar paramagnetic response from the multijunction loops in our samples. We note that these results confirm recent measurements of PME in a niobium disk.[25]

Our results can also be directly related to the PME measured in high-$T_C$ granular superconductors. Returning to the summary of high-$T_C$ experimental data in the Introduction, we have the following.

(a) PME occurs in low FC experiments because flux quanta get trapped in the voids between the grains. In our experiment, this happens for $h_{ac} > 50$ mOe, in the reentrant regime, corresponding to states in upper overlapping branches of Fig. 7. These branches are paramagnetic at small values of the field, and become diamagnetic at higher values, explaining the observed crossover.

(b) PME occurs only if there are weak links between the grains. This follows naturally from our multijunction loop model.

(c) PME appears for strongly coupled grains because high values of $\beta_L$ are required to get hysteretic $\Phi_{TOT}$ vs $\Phi_{EXT}$ curves (see Fig. 7).

The diamagnetic response measured from these materials at small values of magnetic field in ZFC experiments can also be explained within the same scenario. In these experimental conditions, most of the loops will be in states corresponding to the diamagnetic branch crossing $\Phi_{TOT}=0$ (see Fig. 7). (In our measurements, this corresponds to the Meissner-like regime.)

Perhaps the most striking discrepancy between our results and the measurements reported for granular superconductors is that granular samples are either paramagnetic or diamagnetic when measured with dc methods, with no reentrance measured in $\chi_{ac}$. By contrast, our samples and simulations show a reentrance in $\chi_{ac}$ as the temperature is lowered. The are two reasons for this. First, a granular system has a distribution of critical currents and loop sizes, and thus a distribution of $\beta_L$'s. (Typical values of $\beta_L$ in granular high-$T_C$ superconductors are in the range 5–300.[26]) Second, although the value for $\beta_C$ is not known for typical grains, it is probably less than one. In our simulations, only loops with large $\beta_L$ and $\beta_C$ display reentrant ac susceptibility. By contrast, to have a paramagnetic $\chi_{dc}$, only a multibranch solution (i.e., high $\beta_L$) is needed; see Fig. 7.

We conclude that the phenomena causing the reentrance we observed in Josephson-junction arrays should also exist in granular superconductors. We expect these phenomena to appear either as PME, in the case of dc susceptibility measurements, or as an anomalous increase of dissipation at low temperature, in the case of ac susceptibility measurements. Numerical simulations of two-dimensional Josephson-junction networks with a distribution of characteristic parameters $\beta_L$ and $\beta_C$ would be very useful for further theoretical investigations of these phenomena.

## ACKNOWLEDGMENTS

We thank M. G. Forrester, A. W. Smith, and C. B. Whan for their technical help in the experiments. We also thank R. Newrock, A. Sanchez, and D. X. Chen for useful discussions. We gratefully acknowledge financial support from the U.S. Air Force Office of Scientific Research through Grant No. F496209810072, and from the National Science Foundation through Grant No. DMR9732800. F.M.A.M. also gratefully acknowledges financial support from the Brazilian Agency FAPESP, under Grant No. 96/7704-6.

---

*Electronic address: paola@squid.umd.edu

†Permanent address: Universidade Federal de São Carlos (UFSCar), Departamento de Fisica, Grupo de Supercondutividade e Magnetismo, São Carlos–São Paulo 13565-905, Brazil. Electronic address: faraujo@squid.umd.edu.